\documentclass[11pt,a4paper]{article}
\usepackage{jheppub}
\usepackage{amsmath}
\usepackage{hyperref}
\usepackage[utf8]{inputenc}
\usepackage[T1]{fontenc}
\usepackage{lmodern} 
\usepackage{amssymb}
\usepackage{mathtools}
\usepackage{mathrsfs}
\usepackage{amsfonts}
\usepackage{bbm}
\usepackage{relsize}
\usepackage{enumerate}
\usepackage{array}
\usepackage{booktabs}
\usepackage{textcomp}
\usepackage{caption}
\usepackage{cite}
\usepackage{cancel}
\usepackage{units}
\usepackage{multirow}
\usepackage{pst-all}
\usepackage{longtable}
\usepackage{relsize}
\usepackage[]{natbib}

\title{About the phase space of SL(3) Black Holes}

\author[a]{Alejandro Cabo-Bizet}
\author[a,c]{V. I. Giraldo-Rivera}

\affiliation[a]{SISSA and INFN,
Via Bonomea 265, 34128 Trieste, Italy}
\affiliation[c]{ICTP,  Strada Costiera 11, 34014 Trieste, Italy}

\emailAdd{acabo@sissa.it}
\emailAdd{vgiraldo@ictp.it}

\begin{document}
\title{About the phase space of SL(3) Black Holes}

\date{\today}
\abstract{
In this note we address some issues of recent interest, related to the asymptotic symmetry algebra of higher spin black holes in $sl(3,\mathbb{R})\times sl(3,\mathbb{R})$ Chern Simons (CS) formulation.
We compute the fixed time Dirac bracket algebra that acts on two different phase spaces. Both of these spaces contain black holes as zero modes.  The result for one of these phase spaces is explicitly shown to be isomorphic to $W^{(2)}_3\times W^{(2)}_3$ in first order perturbation theory.}
 
\arxivnumber{}

\maketitle

\flushbottom
\section{Introduction}
Higher spin theories \citep{Fradkin:1987ks,Prokushkin:1998bq,Vasiliev:1999ba, Henneaux:2010xg,Campoleoni:2010zq}
in 3D, have been of great interest recently and
specifically, the study of higher spin black holes in the Chern-Simons formulation has been one of the most active lines of research 
\citep{Gutperle:2011kf, Ammon:2011ua, Kraus:2012uf, Banados:2012ue, David:2012iu, Ferlaino:2013vga, Perez:2014pya, Perez:2012cf, 
Compere:2013gja, Gutperle:2013oxa, deBoer:2013gz, Datta:2013qja, Bunster:2014mua}. 

The 3D Chern-Simons (CS) is a theory of pure gauge degrees of freedom. However, in backgrounds with conformal boundaries, like $AdS_3$, it is not a trivial theory. To have a well defined variational principle, boundary terms should be added to the original action. These boundary terms are designed to make the total action stationary under motion in a given region of the moduli space of flat connections.  The selection of that region, a.k.a. imposition of boundary conditions, defines the domain of the moduli space to work with: the phase space. Motion outside of the phase space does not leave the action invariant and it is incompatible with the variational principle. The corresponding gauge transformations we will call from now on ``non residual''.  Motion inside the phase space instead, leaves the total action invariant by construction, then it is admissible. The corresponding gauge transformations we will call from now on ``residual'' and some of them (these are called improper) emerge as global symmetry transformations \citep{Benguria:1976in}. It is very important to stress that throughout this paper we will use the term phase space in the sense stated above, and not to denote all possible initial data in a given Cauchy surface, as it is usually done \footnote{We should stress that this classification (residual, non residual) should not be confused with the usual (proper, improper) \citep{Benguria:1976in, Regge:1974zd, Campoleoni:2011hg}. The latter being applied onto residual gauge transformations (those that preserve the gauge fixing and boundary conditions). For example, improper, are those gauge transformations that change the near boundary behaviour while being residual. We thank 
a referee for drawing our attention to the importance of stressing this point. }.

In the last few years some families of phase spaces have been argued to contain generalisations of the BTZ black hole \citep{Banados:1992wn}: They are called higher spin black holes. See \citep{Gutperle:2011kf, Bunster:2014mua}. Each one of these families is labeled by a set of numbers $\mu,\bar{\mu}$ usually called  chemical potentials. The name deriving from the fact that they can be identified with the chemical potentials of conserved higher spin currents in a 2D CFT.   Recently, attention has been paid to the fixed time canonical bracket structure of these families \citep{Compere:2013gja, Bunster:2014mua} (studies for highest weight boundary conditions can be found at \citep{Campoleoni:2010zq,Campoleoni:2011hg,Henneaux:2010xg}).  One main point of interest regards the classification of charges of generalised black hole solutions. 
 In this note we will address issues related to this last point. We will do it in a perturbative framework and for the case in which the gauge algebra is $sl(3,\mathbb{R})$, in order to be able to explicitly compute Dirac brackets.

The outline of the paper is as follows. In section \ref{sec:ASA} we review the Regge-Teitelboim (RT) formalism in the framework of CS theories in 3D spacetime with boundaries.  In section \ref{DiracBPrincipal} we compute explicitly the fixed time Dirac bracket algebra associated to a phase space \citep{Compere:2013gja}, that in order to avoid confusion afterwards, we denote as $P$-phase space. In section \ref{subsec:RT} we compute the same algebra but by using the method of variation of generators. We have checked that this algebra is not isomorphic to $W_3$. In sections \ref{redW3} and \ref{RTW3}, we provide an interpretation of a related result presented in \citep{Compere:2013gja}. Our interpretation is consistent with the conclusions given in \citep{Bunster:2014mua}. In section \ref{DiracBDiagonal} we compute the fixed time Dirac bracket algebra acting on a different phase space, that again to avoid confusion with the previous $P$-phase space, we denote as $D$- phase space.  This phase space contains black holes as well, and we will show that its fixed time Dirac bracket structure is isomorphic to $W^{(2)}_3$ \citep{Bunster:2014mua
}, up to first order in perturbations of the inverse of the chemical potential $\nu_3$.

\section{The Regge-Teitelboim formalism in 3D CS with boundaries} \label{sec:ASA}

We start this section by reviewing the Regge-Teitelboim (RT) formalism in the context of Chern Simons theory in a 3D space with boundaries. Firstly, we provide some tips that the reader should keep in mind for the rest of the paper.
\begin{itemize}
\item Along our discussion we will use the $\lambda=3$ truncation of  $hs(\lambda)$ to $sl(3,\mathbb{R})$. However many of the procedures to be reviewed in the next section do generalise straightforwardly to any of the truncations gotten for positive integer  $\lambda$. 
\item The super index $(0)$ in a given quantity $X$ stands for its restriction to the Cauchy surface $X^{(0)}$. Or equivalently to its initial condition under a given flow equation.
\item The symbol $\delta$ stands for an arbitrary functional variation whereas $\delta_\Lambda$ stands for a variation due to a residual gauge transformation $\Lambda$.
\end{itemize}

Let us denote by $(\mathcal{A},\bar{\mathcal{A}})$ the left and right $sl(3, \mathbb{R})$-valued connections of interest. Let us focus on  the sector $\mathcal{A}$ and let us denote the space-time coordinates by $(\rho, x_1, x_2)$. The Chern Simons action supplemented by a boundary term is
\begin{equation}
S_{CS}=\int tr\left(\mathcal{A}d\mathcal{A}+\frac{2}{3}\mathcal{A}^3\right)+I_{bdry}.
\end{equation}
Part of the $hs(\lambda)$ \footnote{See appendix \ref{app:Conv} for notations, conventions and definitions concerning the $hs(\lambda)$ algebra.} gauge freedom is fixed  by the choice
 \begin{equation}\mathcal{A}_\rho=V^{2}_0, ~\left( \bar{\mathcal{A}}_\rho=-V^{2}_0\right). \label{Arho1}\end{equation}
The $(1,\rho)$ and $(2,\rho)$ components of the equations of motion $d\mathcal{A}+\mathcal{A}^2=0$ impose the form
\begin{equation}
\mathcal{A}_{a}=b A_{a} b^{-1}, ~  \, b=e^{-\rho V^{2}_0}~\bigg( \bar{\mathcal{A}}_{a}=\bar{b} A_{a} \bar{b}^{-1}, ~  \, \bar{b}=e^{\rho V^{2}_0}\bigg),\label{conjugation}
\end{equation}
with $a=1,2$\footnote{ From now on we will focus on the unbarred sector $\mathcal{A}$. The results for the barred sector $\bar{\mathcal{A}}$ can be obtained in the same way.}. The remaining $(1,2)$ components read
\begin{equation}
d A+A^2=0, ~ d\equiv dx^a \partial_a. \label{eom}
\end{equation}
Up to this point we have twice as many variables than equations. Equation \eqref{eom} can be thought of as:
\begin{itemize}
\item $x_2$ evolution equation for $A_1$. ~($\partial_2 A_1+\ldots=0$).
\end{itemize}
where the $\ldots$ define quantities that do not involve derivatives with respect to $x_2$.\\

From this point of view $A_2$ is an arbitrary source and the Cauchy surface initial condition is $A_1|_{x_2=fixed}$. The arbitrariness of the source $A_2$ represents an extra gauge freedom that tunes the $x_2$  evolution of a Cauchy data surface $A_1|_{x_2=fixed}$. Should we make the choice $A_2=0$, evolution is trivial and all Cauchy surfaces have the same data $A_1(x_1)$. Data $A_1(x_1)$ and $A_1(x_1)+\delta_\Lambda A_1(x_1)$ are physically inequivalent as the gauge degeneracy has been already fixed.

However, notice that one can map $\delta_\Lambda A_1(x_1)$ to an ``improper" $hs(\lambda)$ residual gauge transformation with parameter $\Lambda(x_1)$\footnote{In terms of the calygraphic components $\mathcal{A}$, the gauge parameter is $b \Lambda(x_1) b^{-1}$, in such a way that it preserves the $hs(\lambda)$ gauge choice $(\mathcal{A}_\rho,\mathcal{A}_2)=(V^2_0,0)$ (and hence it represents a ``residual'' gauge transformation). The gauge transformation $\Lambda$, while preserving the gauge choice and hence  being ``residual'', is usually called ``improper" due to the fact that it changes the near boundary data, namely it defines motion in the physical phase space. In a manner that will be explicitly shown below, these transformations define global symmetries. }. In this way the gauge choice $A_2=0$ is preserved and
\begin{equation}
\delta_\Lambda A_1(x_1)\equiv\partial_1 \Lambda(x_1)+[A_1,\Lambda].
\end{equation}
The gauge parameters $\Lambda$ carry thence some physical meaning, they will define global charges $Q(\Lambda)$ whose Poisson bracket with the initial data $A_1(x_1)$ will generate the changes $\delta A_1(x_1)$.  In fact, in virtue of what was said, it results that
\begin{equation}
Q(\Lambda)= G|_{\mathcal{A}_\rho=V^2_0,\mathcal{A}_1=b A_1 b^{-1}}(b \Lambda(x_1) b^{-1}). \label{QLambda}
\end{equation}
Where $G$ is the generator of gauge transformations in a given Cauchy surface before imposing any second class constraint. Even though we did not make it explicit in \eqref{QLambda}, we have also imposed $\mathcal{A}_2=0$.

Before defining $G$ let us stress that in the following paragraph we do not impose neither \eqref{Arho1} nor \eqref{conjugation} which are not compatible (namely, there are second class constraints) with the $x_2=fixed$ Poisson bracket algebra
\begin{equation}
\{\mathcal{A}_1,\mathcal{A}_\rho\}_{PB}=-\{\mathcal{A}_\rho,\mathcal{A}_1\}_{PB}=V^1_0\delta^{(2)}.\label{pbinit}
\end{equation}
Where by $V^1_0$ we mean the identity operator in the $hs(\lambda)$ algebra (See appendix \ref{app:Conv}). However we are free to take $\mathcal{A}_2=0$ as it is compatible (first class) with \eqref{pbinit}.
The quantity
\begin{equation}
G(\Gamma)\equiv\int dx_1tr(\Gamma \mathcal{A}_1)|_{\rho=\infty}+\int d x_1 d\rho ~tr(\Gamma \mathcal{F}_{1\rho}), \label{GGamma}
\end{equation}
is defined over each $x_2=fixed$ Cauchy surface and obeys the following properties
\begin{eqnarray}
\{G{(\Gamma}),\mathcal{A}_{1,\rho}\}_{PB}&=&D_{1,\rho}\Gamma \equiv \delta_{\Gamma} \mathcal{A}_{1,\rho},\nonumber\\
\delta_{\mathcal{A}_{1}}G{(\Gamma)}&=&-\int dx_1 d\rho ~ tr \left(D_{\rho}\Gamma\delta \mathcal{A}_{1}\right), \label{Ggene}
\end{eqnarray}
under the brackets \eqref{pbinit}. Namely, it generates the gauge transformations on a given Cauchy surface under \eqref{pbinit}, and it is properly differentiable under off-shell variations $\delta \mathcal{A}_1$.  By computing the gauge variation of \eqref{GGamma} and regrouping some terms one arrives to the algebra
\begin{equation}
\{G(\Gamma_1),G(\Gamma_2)\}_{PB}\equiv\delta_{\Gamma_1} G(\Gamma_2)=G([\Gamma_1,\Gamma_2])-\int dx_1 ~tr(\Gamma_1 \partial_1\Gamma_2), \label{algebraG}
\end{equation}
which is inherited through \eqref{QLambda} by the $Q(\Lambda)$'s.

 In fact, after plugging \eqref{GGamma} into \eqref{QLambda} one gets
\begin{equation}
Q(\Lambda)=\int dx_1tr(\Lambda A_1).
\end{equation}
From the first line in \eqref{Ggene} and after imposing the second class constraints \eqref{Arho1} and \eqref{conjugation} we arrive to
\begin{equation}
\{Q(\Lambda),A_1\}_{PB}=D_1 \Lambda\equiv \delta_\Lambda A_1(x_1) ,\label{rstra}
\end{equation}
which after taking $\Lambda=\delta^{2} \tau_a$, $A_1=A^b_1 \tau_b$ reduces to the Kac-Moody algebra
\begin{eqnarray}
\{A_1^a(x_1),A_1^b(y_1)\}_{PB}= f^{ab}_{~~c}A_1^c \delta(x_1-y_1)-g^{ab}\partial_{x_1}\delta(x_1-y_1),\label{KMAlgebra}
\end{eqnarray}
where $g^{ab}$ is the inverse of the Killing metric, $g_{ab}=tr \left(\tau_a \tau_b\right)$, that is also used to raise indices. To lower indices we use the Killing metric $g_{ab}$ itself. For instance $f^{ab}_c=g^{a\bar{a}}g^{b\bar{b}}g_{c\bar{c}} f_{\bar{a}\bar{b}}^{~~\bar{c}}$. Where $[\tau_a,\tau_b]=f_{a b}^{~~c}\tau_c$. Notice that the same result (\ref{KMAlgebra}) can be deduced from (\ref{algebraG}) and the definition (\ref{QLambda}).

It is worth to notice that in the previous definition of $G$, the gauge parameter $\Gamma$ was supposed to be field independent. Should this not be the case, then (\ref{GGamma}) should be replaced by
\begin{equation}
G(\Gamma)\equiv B(\Gamma,\mathcal{A})+\int d x_1 d\rho ~tr(\Gamma \mathcal{F}_{1\rho}),\label{genNonLinear}
\end{equation}
where the boundary term $B$ is such that
\begin{equation}
\delta_{\mathcal{A}_1}B(\Gamma)=\int dx_1 tr(\Gamma \delta\mathcal{A}_1)|_{\rho=\infty}.\label{boundarytNonLinear}
\end{equation}
Is easy to check that (\ref{genNonLinear}) still obeys the properties (\ref{Ggene}), but in a weak sense, namely up to terms that vanish when one imposes the equations of motion, $\mathcal{F}_{1\rho}=0$. Clearly when $\Gamma$ is field independent both definitions (\ref{GGamma}) and (\ref{genNonLinear}) are equivalent. But  (\ref{genNonLinear}) is more general. So we will stick to \eqref{genNonLinear}.

 For later use we impose (\ref{Arho1}), (\ref{conjugation}), and $\Gamma=b\Lambda b^{-1}$, onto (\ref{boundarytNonLinear}) and rewrite it  as
\begin{equation}
\delta Q(\Lambda)=\int dx_1 tr(\Lambda \delta A_1).\label{boundarytNonLinear2}
\end{equation}
Where now we note that the $\rho$ dependence has disappeared, and the non linearity of $\Gamma$ is inherited by $\Lambda$. The integration of (\ref{boundarytNonLinear2}),  $Q(\Lambda)$, generates the residual gauge transformations that preserve any further constraint,  with $\Lambda$ being the corresponding residual gauge parameter. From (\ref{rstra})  we have then a way to find out the Poisson brackets on  a further reduced phase space.

A shortcut to find out the algebra without integrating (\ref{boundarytNonLinear}) is at hand. After use of the equivalence relation in (\ref{algebraG}) inherited by the $Q$, together with (\ref{boundarytNonLinear2}) one gets
 \begin{equation}
 \{Q(\Lambda_1),Q(\Lambda_2)\}_{PB}\equiv\delta_{\Lambda_1} Q(\Lambda_2)=-\int dx_1~ tr(\Lambda_1 D_1 \Lambda_2). \label{CovPB}
 \end{equation}
 In this way we just need to use $A_1$ and the residual gauge parameter $\Lambda$ to evaluate the RHS \citep{Compere:2013gja}. We will not resort to this way.

 Notice also, that in the process we have been neglecting total derivative terms with respect to $x_1$ under integration. To take care of them, one imposes boundary conditions on the field and gauge parameters, like for instance periodicity under $x_1 \rightarrow x_1+2\pi$. In the next section we will study a case in which such a periodicity is lost due to the use of perturbation theory. 
 



\section{Two phase spaces of $sl(3,\mathbb{R})$ black holes. }

In this section we go on to analyse the phase space of $sl(3,\mathbb{R})$ CS theories with modified boundary condition. By modified we mean others than the highest weight condition used in \citep{Henneaux:2010xg,Campoleoni:2010zq}. With that goal in mind, we 
compute explicitly the Dirac bracket algebra with the Dirichlet boundary conditions introduced in \citep{Gutperle:2011kf} and studied in \citep{Compere:2013gja}.  In subsection \ref{DiracBPrincipal} we compute the fixed time Dirac bracket algebra that comes from the imposition of 6 constraints onto the $sl(3,\mathbb{R})$ Kac Moody algebra (\ref{KMAlgebra}). In section \ref{subsec:RT} we recompute the same bracket algebra by use of the method of variation of the generators that was used in  section \ref{sec:ASA} to compute the Kac Moody algebra (\ref{KMAlgebra}). Let us be more precise in summarising this last result. The bracket algebra obtained by the method of variation of generators will depend on a set of integration constants that describe all possible field redefinitions of the smearing gauge parameter. As will be checked in subsection \ref{subsec:RT}, for a specific choice of these integration constants this algebra will coincide with the Dirac bracket algebra reported in section \ref{DiracBPrincipal}.  

Additionally, we must say, that there is another choice of the aforementioned integration constants that, as shown in section \ref{redW3}, define a $W_3$ bracket algebra (up to 
redefinitions of the generators). In subsection \ref{RTW3} we check that such a choice of integration constants is equivalent to performing a non residual gauge transformation to the highest weight choice \citep{Henneaux:2010xg,Campoleoni:2010zq}. This is also the redefinition used by the authors in \citep{Compere:2013gja} to arrive to a $W_3$ symmetry transformation
. Let us be more specific before entering in details. As already said and shown in subsections \ref{redW3} and \ref{RTW3}, this choice of integration constants consists of both, a redefinition of the residual gauge transformation parameters and a redefinition of the phase space parameters (the background connection). The field dependent redefinition of the residual gauge parameters to be used in this case differs with the one used in the case mentioned in the previous paragraph. This difference suggests, and we will check so, that the Dirac bracket algebra we have referred to in the last sentence of the previous paragraph is not isomorphic to $W_3$ \citep{Bunster:2014mua}. Accordingly, the $W_3$ symmetry transformation, that the authors in \citep{Compere:2013gja} arrive to, after performing the corresponding transformations, is not acting onto the original phase space of parameters (up to coordinates redefinitions) but onto a different phase space given by the highest weight gauge choice \citep{Henneaux:2010xg,Campoleoni:2010zq}. This last statement will be checked in section \ref{redW3}.

In subsection \ref{DiracBDiagonal} we consider a different reduction of the $sl(3,\mathbb{R})$ phase space. In this case we classify the $sl(3,\mathbb{R})$ generators according to a diagonally embedded gravitational $sl(2,\mathbb{R})$ and impose less amount of constraints, in total 4, onto the $sl(3,\mathbb{R})$ Kac Moody algebra (\ref{KMAlgebra}). By explicit computation the fixed time Dirac bracket algebra in this new phase space, is shown to be isomorphic to $W^{(2)}_3$ up to first order in perturbations of the inverse of the chemical potential $\nu_3$.

\subsection{Explicit computation of Dirac bracket algebra in $P$-phase space} \label{DiracBPrincipal}

We will impose 6 second class constraints (boundary conditions) onto the phase space $\eqref{KMAlgebra}$ of 3D CS theory with Lie algebra $sl(3,\mathbb{R})$. The reduced phase space will be called $P$-phase space.  Specifically, we compute the Dirac bracket  algebra on the reduced phase space, in a Cauchy surface at fixed $t_0$. The main point of this section is to show by explicit computation that this algebra is not isomorphic to the $W_3$ algebra.
 
 We start by defining what we call $P$-phase space
. First we relax the condition $A_2=0$ used in section \ref{sec:ASA}. Besides (\ref{Arho1}) and (\ref{conjugation}), we impose the following constraints
\begin{eqnarray}\nonumber
A_1&=&V^{2}_1+\mathcal{L}V^{2}_{-1}+\mathcal{W}V^{3}_{-2},
\\
A_2&=&\mu_3 \left(V^{3}_{2}
+\text{ lower components } \right),\label{Dirichlet1}
\end{eqnarray}
where the highest weight elements $\left(\mathcal{L},\mathcal{W},\ldots\right)$ are arbitrary functions of $(x_1,x_2)$. From now on to save some notation we denote the set of all of them $\left(\mathcal{L},\mathcal{W},\ldots\right)$ as $\mathcal{M}$. 
The boundary conditions that define the phase space of connections of the form \eqref{Dirichlet1}(that we call from no on, $P$-phase space), were introduced in \citep{Gutperle:2011kf, Compere:2013gja}.

To completely precise (\ref{Dirichlet1}), flatness conditions must be imposed. The flatness conditions along the generators $V^{s}_{m_s \geq -s+1}$ provide algebraic equations for the ``lower components" in terms of $\left(\mathcal{M},\partial_2 \mathcal{M}\right)$. 
\begin{eqnarray}
A_2=\mu_3 \left(V^{3}_2+2 \mathcal{L}V^{3}_0-\frac{2}{3}\partial_1\mathcal{L}V^{3}_{-1} +\left(\mathcal{L}^2+\frac{1}{6}\partial^2_1\mathcal{L}\right)V^{3}_{-2}-2 \mathcal{W}V^{2}_{-1}\right).\label{A2vero}
\end{eqnarray}
 The remaining ones provide the $x_2$-flow equations
 \begin{eqnarray}
 \partial_2 \mathcal{L}=-2 \mu_3 \partial_1 \mathcal{W},~~ \partial_2 \mathcal{W}= \mu_3\left(\frac{8}{3} \mathcal{L}\partial_1\mathcal{L}+\frac{1}{6}\partial_1^3\mathcal{L}\right),\label{x2equations}
 \end{eqnarray}
which determine the $\mathcal{M}$ out of the initial conditions $\mathcal{M}(x_1,0)$.  Solutions can be found in terms of perturbations of the chemical potential $\mu_3$ and will have the generic form
\begin{equation}
\mathcal{M}=\mathcal{M}^{(0)}+\mu_3  \left(x_2 \mathcal{M}^{(1)}+\mathcal{M}^{(0)}_1\right)
+O(\mu_3^2), \label{patronM}
\end{equation}
where $\mathcal{M}^{(1)}$, 
are local functionals of the initial conditions $\mathcal{M}^{(0)}$, $\mathcal{M}^{(0)}_1$. 
Notice that the integration constants $\mathcal{M}^{(0)}_1$ 
are just shifts in $\mathcal{M}^{(0)}$. In general we will take $\mathcal{M}^{(0)}_1$ as the most general functional of $x_1$ and $\mathcal{M}^{(0)}$ consistent with dimensional analysis. The explicit dependence in $x_1$ will play an important role. 
 
  To make things easier we start by computing the brackets on a Cauchy surface at fixed $x_2$. In this case the phase space is given by the $sl(3,\mathbb{R})$ valued function of $x_1$ that defines the $x_1$ component $A_1$ in \eqref{Dirichlet1}. 
  
   Let a generic $sl(3,\mathbb{R})$ valued function of $x_1$ be
 \begin{eqnarray}
 a(x_1)=A^{s}_{m_s} V^{s}_{m_s}=A^a V_a,~~~~~~~~~~~~~~~~~
\nonumber\\ V_a=\left(V^{2}_1,V^{2}_0,V^{2}_{-1},V^{3}_2,V^{3}_1,V^{3}_0,V^{3}_{-1},V^{3}_{-2}\right).\label{Ppalemb}
 \end{eqnarray}
We start from the Kac-Moody algebra (\ref{KMAlgebra}) and proceed to impose the following 6 second class constraints
 \begin{equation}
 C^i=\bigg(A^{2}_1-1,A^{2}_0,A^{3}_2,A^{3}_1,A^{3}_0,A^{3}_{-1}\bigg), \label{ConstraintsW3}
 \end{equation}
onto $a(x_1)$, but first we choose the integration constants $\mathcal{M}^{(0)}_1$ to be
\begin{eqnarray}
\mathcal{L}^{(0)}_1&=& 2 \mathcal{W}^{(0)} +2 x_1 \partial_1 \mathcal{W}^{(0)},\nonumber\\
\mathcal{W}^{(0)}_1&=&-{\mathcal{L}^{(0)}}^2-\frac{1}{6}\partial^2_1 \mathcal{L}^{(0)}-x_1\frac{1}{6}\left(16 \mathcal{L}^{(0)}\partial_1 \mathcal{L}^{(0)}+\partial_1^3 \mathcal{L}^{(0)}\right),\label{DiracChoice}\end{eqnarray}
 From now on, to save space we will not write down the explicit $t_0$ dependence but the reader should keep in mind that the full result is recovered by making the substitutions
 \begin{eqnarray}
 \mathcal{L}^{(0)}&\rightarrow&\mathcal{L}^{(0)}+\mu_3 t_0 \mathcal{W}^{(0)}+O(\mu_3^2),\nonumber\\
 \mathcal{W}^{(0)}&\rightarrow&\mathcal{W}^{(0)}+\mu_3 t_0\frac{1}{12}\left(16 \mathcal{L}^{(0)}\partial_1
\mathcal{L}^{(0)}+\partial_1^3 \mathcal{L}^{(0)}\right)+O(\mu_3^2), \label{subs}
 \end{eqnarray}  at the very end. 
 
 The constraints (\ref{ConstraintsW3}) define the Dirac bracket 
\begin{equation}
\{A^a(x_1),A^b(y_1)\}_D=\{A^a(x_1),A^b(y_1)\}_{PB}-\left(\{A^a,C^i\}_{PB} M_{ij}\{C^j ,A^b\}_{PB}\right)(x_1,y_1),\label{DBras}
\end{equation}
in the reduced phase space with configurations $A^a=(\mathcal{L}^{(0)},\mathcal{W}^{(0)})$.

 The object $M_{ij}(x_1,y_1)$ is the inverse operator of $\{C^i(x_1),C^j(x_2)\}_{PB}$, whose non trivial components are computed to be
\begin{eqnarray}
M_{12}=\frac{1}{2}\delta_{x_1y_1},~ M_{21}=-M_{12},~ M_{22}=\frac{1}{2}\partial_{x_1}\delta_{x_1y_1}, ~M_{36}=-\frac{1}{4}\delta_{x_1y_1},~~~~~~~~~~~~~~\nonumber\\
M_{45}=\frac{1}{12} \delta_{x_1y_1},~M_{46}=-\frac{1}{12} \partial_{x_1}\delta_{x_1y_1},~M_{54}=-M_{45} , ~ M_{55}=\frac{1}{24}\partial_{x_1} \delta_{x_1y_1}, ~~~~~~~~~~\nonumber\\
M_{56}=-\frac{1}{4}(\mathcal{L}^{(0)} \delta_{x_1y_1}+\frac{1}{6}\partial_{x_1}^2 \delta_{x_1y_1}),~M_{63}=-M_{36},~M_{64}=M_{46},~ M_{65}=-M_{56},~~~~~~~~~~~~~\nonumber\\~M_{66}=-\frac{1}{4}\left(\partial_{x_1}\mathcal{L}^{(0)}\delta_{x_1y_1}+2 \mathcal{L}^{(0)}\partial_{x_1}\delta_{x_1y_1}+\frac{1}{6}\partial_{x_1}^3\delta_{x_1y_1}\right).~~~~~~~~~~~~~~~~~~~~~~~ \label{M0}
\end{eqnarray}
It is easy to check that $M_{ij}(x_1,y_1)=-M_{ji}(y_1,x_1)$ as it should be.
After some algebra (\ref{DBras}) takes the explicit form
\begin{eqnarray}
\{\mathcal{L}^{(0)}(y_1),\mathcal{L}^{(0)}(x_1)\}_D&=&\partial_{x_1}\mathcal{L}^{(0)} \delta_{x_1y_1}+2 \mathcal{L}^{(0)} \partial_{x_1}\delta_{x_1y_1}+\frac{1}{2}\partial_{x_1}^3\delta_{x_1y_1},\nonumber\\
\{\mathcal{L}^{(0)}(y_1),\mathcal{W}^{(0)}(x_1)\}_D&=&2\partial_{x_1}\mathcal{W}^{(0)} \delta_{x_1y_1}+3\mathcal{W}^{(0)} \partial_{x_1}\delta_{x_1y_1},\nonumber\\
\{\mathcal{W}^{(0)}(y_1),\mathcal{W}^{(0)}(x_1)\}_D&=&-\frac{1}{6} \left(16 \mathcal{L}^{(0)} \partial_{x_1} \mathcal{L}^{(0)}+ \partial^3_{x_1} \mathcal{L}^{(0)}\right) \delta_{x_1y_1} -\nonumber\\&&\frac{1}{12} \left(9 \partial^2_{x_1} \mathcal{L}^{(0)}+32 {\mathcal{L}^{(0)}}^2\right) \partial_{x_1} \delta_{x_1y_1}-\frac{5}{4} \partial_{x_1} \mathcal{L}^{(0)} \partial^2_{x_1} \delta_{x_1y_1}-\nonumber\\&&~~~~~~~~~~~~~~~~~~~~~~~~\frac{5}{6} \mathcal{L}^{(0)} \partial^3_{x_1} \delta_{x_1y_1}-\frac{1}{24} \partial^5_{x_1} \delta_{x_1y_1},\label{ASA2}
\end{eqnarray}
where all the $\mathcal{L}^{(0)}$ and $\mathcal{W}^{(0)}$ in the right hand side are evaluated on $x_1$. The brackets \eqref{ASA2}, define a $W_3$ algebra at fixed light cone coordinate $x_2$ slices\footnote{This is, when evolution along $x_2$ is considered.} for the phase space \eqref{Dirichlet1} \citep{Compere:2013gja,deBoer:2014fra}. Notice that in this case,  the $\mu_3$ dependence is implicit in the fields through the redefinitions (\ref{subs}).

Now we go a step forward to compute the Dirac bracket on a Cauchy surface at fixed time $t_0$. This time the constraints will look like
\begin{equation}
 C^i=\left(A^{2}_1-1,A^{2}_0,A^{3}_2- \mu_3,A^{3}_1,A^{3}_0-2\mu_3\mathcal{L},A^{3}_{-1}+\frac{2}{3}\mu_3 \partial_1 \mathcal{L}\right), \label{Constmu3}
 \end{equation}
and the corresponding first order in $\mu_3$ corrections to (\ref{M0}) are
\begin{eqnarray}
M^1_{14}=\frac{1}{6}\delta_{x_1y_1}, ~M^1_{15}=-\frac{1}{6}\partial_{x_1}\delta_{x_1 y_1}, ~ M^1_{16}=\delta_{x_1y_1}\mathcal{L}^{(0)}+\frac{1}{4}\partial_{x_1}^2\delta_{x_1 y_1},~~~~~~~~~~\nonumber\\
M^1_{23}=-\frac{1}{2}\delta_{x_1y_1}, ~M^1_{24}=\frac{1}{3}\partial_{x_1}\delta_{x_1y_1},~M^1_{25}=-\frac{2}{3}\delta_{x_1 y_1}\mathcal{L}^{(0)}-\frac{1}{4}\partial_{x_1}^2\delta_{x_1 y_1},~~~~~~~~\nonumber\\
M^1_{26}=\frac{5}{3}\delta_{x_1y_1}\partial_{x_1} \mathcal{L}^{(0)}+\frac{7}{3}\partial_{x_1}\delta_{x_1 y_1}\mathcal{L}^{(0)}+\frac{1}{3}\partial_{x_1}^3 \delta_{x_1y_1}, ~ M^1_{32}=-M^1_{23}, M^1_{41}=-M^1_{14},~~~~~~~~\nonumber\\  M^1_{42}=M^1_{24}, ~M^1_{51}=M^1_{15}, ~ M^1_{52}=-M^1_{25},~M^1_{56}=-\frac{1}{6}\delta_{x_1y_1} \mathcal{W}^{(0)}, ~ M^1_{61}=-M^{1}_{16},  ~~~~~~~\nonumber\\M^{1}_{62}=\frac{2}{3} \delta_{x_1y_1}\partial_{x_1}\mathcal{L}^{(0)}+ \frac{7}{3}\partial_{x_1}\delta_{x_1y_1}\mathcal{L}^{(0)}+\frac{1}{3}\partial_{x_1}^3 \delta_{x_1y_1},~M^1_{65}=-M^1_{56},~~~~~~~~~~~~~\nonumber\\ M^1_{66}=-\frac{1}{3}\delta_{x_1y_1} \partial_{x_1}\mathcal{W}^{(0)}-\frac{2}{3}\partial_{x_1}\delta_{x_1y_1} \mathcal{W}^{(0)}. ~~~~~~~~~~~~~~~~~~~~~~~~~~ \label{M1}
\end{eqnarray}
Again it is easy to check that $M^1_{ij}(x_1,y_1)=-M^1_{ji}(y_1,x_1)$. From  (\ref{DBras}), (\ref{M0}) and (\ref{M1}) we compute the corresponding Dirac bracket. They can be checked to obey the compatibility property $\{C^i,\ldots\}_D=0$. 

The corrections to (\ref{ASA2}) are given by
\begin{eqnarray}
\{\mathcal{L}^{(0)}(y_1),\mathcal{L}^{(0)}(x_1)\}_D&=&\ldots+2 \mu_3 \partial_{x_1} \mathcal{W}^{(0)} \delta_{x_1y_1}+4 \mu_3 \mathcal{W}^{(0)} \partial_{x_1}\delta_{x_1y_1}, \nonumber \\ \{\mathcal{L}^{(0)}(y_1),\mathcal{W}^{(0)}(x_1)\}_D&=&\ldots -\mu_3\left(\frac{8}{3}\mathcal{L}^{(0)}\partial_{x_1}\mathcal{L}^{(0)} \delta_{x_1y_1}+\frac{1}{6} \partial_{x_1}^3\mathcal{L}^{(0)}\delta_{x_1y_1}+ \right.\nonumber\\&&~~~~~~~~~~
\left. \frac{13}{3}\mathcal{L}^2  \partial_{x_1}\delta_{x_1y_1}+\frac{4}{3}\partial_{x_1}^2\mathcal{L}^{(0)}\partial_{x_1}\delta_{x_1y_1}+\right. \nonumber\\&&\ \left.~~~~~~\frac{25}{6}\partial_{x_1}\mathcal{L}^{(0)} \partial_{x_1}^2\delta_{x_1y_1}+\frac{11}{3} \mathcal{L}^{(0)}\partial_{x_1}^3 \delta_{x_1y_1}+\frac{1}{3}\partial_{x_1}^5 \delta_{x_1y_1} \right),\nonumber\\
\{\mathcal{W}^{(0)}(y_1),\mathcal{W}^{(0)}(x_1)\}_D&=&  \ldots -\mu_3\left(\frac{22}{3} \partial_{x_1}(\mathcal{W}^{(0)}\mathcal{L}^{(0)})\delta_{x_1y_1}+\frac{44}{3}\mathcal{L}^{(0)}\mathcal{W}^{(0)}\partial_{x_1}\delta_{x_1y_1}+\right.\nonumber\\ &&~~~~~~~~~~~~~~~~~~~~ \left. \partial_{x_1}^3 \mathcal{W}^{(0)}\delta_{x_1y_1}+\frac{10}{3}\partial_{x_1}^2\mathcal{W}^{(0)}\partial_{x_1}\delta_{x_1y_1}+\right. \nonumber\\&& \left.~~~~~~~~~~~~~~~~~~~~~~4\partial_{x_1}\mathcal{W}^{(0)}\partial_{x_1}^2\delta_{x_1y_1}+\frac{8}{3}\mathcal{W}^{(0)}\partial_{x_1}^3\delta_{x_1y_1}\right), \nonumber \\ \label{Dbras3}
\end{eqnarray}
and can not be reabsorbed by a general analytical redefinition at first order in $\mu_3$
\begin{equation}
\mathcal{L}\rightarrow \mathcal{L}+\mu_3 {\mathcal{L}^0_1}_{hom}, ~ \mathcal{W}\rightarrow \mathcal{W}+\mu_3 {\mathcal{W}^0_1}_{hom},
\end{equation}
where the $({\mathcal{L}^{(0)}_1}_{hom},{\mathcal{W}^{(0)}_1}_{hom})$ are given in the first line of (\ref{Ansatz}). So the fixed time Dirac bracket algebra (\ref{Dbras3}) on the phase space (\ref{Dirichlet1})
is not isomorphic to $W_3$. However as we will see (\ref{Dirichlet1}) can be embedded in a larger phase space whose constrained algebra at fixed time slices will be shown to be isomorphic to $W^{(2)}_3$.\\ 

\subsection{Dirac bracket algebra in the $P$ - phase space: The method of variation of generators} \label{subsec:RT}

For completeness we will recompute the Dirac bracket algebra (\ref{Dbras3}) by use of the method of smeared variation of generators used in the computation of (\ref{KMAlgebra}) in section \ref{sec:ASA}. 


We start by determining the set of residual (and improper) gauge transformations that map the $P$-phase space onto itself, namely, that preserve the set of boundary conditions defining the $P$-phase space. We ask now for the set of linear gauge transformations  preserving the boundary conditions (\ref{Dirichlet1})
\begin{eqnarray}
\delta A_a &=& \partial_{x_a}\Lambda+[A_a,\Lambda] , \label{gaugeCompere} \\
\Lambda&=&\epsilon V^{2}_{1} + \eta V^{3}_{2}+ \text{ higher components},
\end{eqnarray} \footnote{Notice that in (\ref{gaugeCompere}) we have used $\delta$ and not $\delta_{\Lambda}$. In fact we use $\delta_\Lambda A$ to denote the solution of the condition (\ref{gaugeCompere}), meanwhile $\delta$ stands for an arbitrary functional variation.}
where the lowest components $\{\epsilon, \eta \}$ are arbitrary functions of $(x_1, x_2)$. We will denote the set of lowest components $\{\epsilon, \eta\}$ by $\Theta$. The projection along the generators $V^{s}_{ m_s > -s+1}$ of the $x_1$ equation in (\ref{gaugeCompere}) solves algebraically for the highest components in terms of the lowest ones $\Theta$:
\begin{eqnarray}
\Lambda(\epsilon,\eta)=\epsilon V^{2}_1-\partial_1 \epsilon V^{2}_0+\left(\mathcal{L}\epsilon-2\mathcal{W} \eta+\frac{1}{2}\partial_1^2\epsilon \right)V^{2}_{-1}+\eta V^{3}_2-\partial_1 \eta V^{3}_1+\nonumber\\\left(2\mathcal{L}\eta+\frac{1}{2}\partial_1^2\eta\right)V^{3}_0-\left(\frac{2}{3}\partial_1\mathcal{L}\eta+\frac{5}{3}\mathcal{L}\partial_1\eta+\frac{1}{6}\partial_1^3 \eta\right)V^{3}_{-1}+~~\nonumber\\\left(\mathcal{W}\epsilon+\mathcal{L}^2\eta+\frac{7}{12}\partial_1\mathcal{L} \partial_1\eta+\frac{1}{6}\partial_1^2\mathcal{L}\eta+\frac{2}{3}\mathcal{L}\partial_1^2\eta+\frac{1}{4}\partial_1^4 \eta\right)V^{3}_{-2}. \label{gaugeParameter}
\end{eqnarray}
Notice that the $A_2$ component (\ref{A2vero}) can be viewed as a residual gauge parameter $\Lambda(0,\mu_3)$. This is of course a reminiscence of its spurious character. 

The remaining $x_1$ equations provide variations of the gauge field parameters $\mathcal{M}(x_1,x_2)$
\begin{eqnarray}
\delta_\Lambda \mathcal{L}&=&\partial_1 \mathcal{L} \epsilon +2 \mathcal{L} \partial_1 \epsilon-2 \partial_1 \mathcal{W} \eta -3 \mathcal{W} \partial_1 \eta+\frac{1}{2} \partial^3_1 \epsilon,\nonumber \\
  \delta_\Lambda \mathcal{W} &=&\partial_1 \mathcal{W} \epsilon +3 \mathcal{W} \partial_1 \epsilon+\frac{1}{6} \left(16 \mathcal{L} \partial_1 \mathcal{L}+ \partial^3_1 \mathcal{L}\right) \eta +\nonumber \\ &&~~~~~~~~\frac{1}{12} \left(9 \partial^2_1 \mathcal{L}+32 \mathcal{L}^2\right) \partial_1 \eta+\frac{5}{4} \partial_1 \mathcal{L} \partial^2_1 \eta+\frac{5}{6} \mathcal{L} \partial^3_1 \eta+\frac{1}{24} \partial^5_1 \eta,\nonumber\\\label{gaugeVARIATIONS}
\end{eqnarray}
From flatness conditions and the Dirichlet boundary condition to impose, it is clear that any other component variation of the gauge fields can be deduced out of these ones. Demanding the lowest weight components $(V^{2}_1, V^{3}_2)$ of the final $A_2$ connection to be fixed, determines the $x_2$-flow equations 
\begin{eqnarray}
\partial_2 \epsilon=-\mu_3\left(\frac{8}{3}\mathcal{L}\partial_1\eta+\frac{1}{6}\partial_1^3\eta\right),~ \partial_2 \eta=2 \mu_3 \partial_1 \epsilon, \label{x2equationsPar}
\end{eqnarray}
which allow to solve for the gauge parameter $\Theta(x_1,x_2)$ in terms of the initial conditions $\Theta(x_1,0)$. Again, solutions can be found in perturbations of the chemical potential $\mu_3$
 \begin{equation}
\Theta=\Theta^{(0)}+\mu_3\left(x_2 \Theta^{(1)}+\Theta^{(0)}_1\right)
+O(\mu_3^2),\label{gaugePar}
\end{equation}
where the $\Theta^{(1)}$, 
are local functionals of the initial conditions $\Theta^{(0)}$. 
 The $\Theta^{(0)}_1$ 
are shifts of $\Theta^{(0)}$ and we will define them as general functionals of $x_1$, $\mathcal{M}^{(0)}$ and $\Theta^{(0)}$ consistent with dimensional analysis, and linear in the $\Theta^{(0)}$.


Let us define our coordinates $x_1=\frac{1}{2}(t_0+ \phi)$,  $ x_2=\frac{1}{2}(-t_0+ \phi)$ and consider time evolution. This choice of coordinates identify (\ref{Dirichlet1}) with the first two lines in equation (3.1) of \citep{Compere:2013gja} under our conventions \footnote{Should we have chosen $x_1=\phi$ and $x_2=t$ the fixed time Dirac bracket algebra of (\ref{Dirichlet1})  is seen to be $W_3$ \citep{Bunster:2014mua}.}.

The Cauchy data at a fixed time slice and the corresponding residual gauge transformations are
\begin{eqnarray}
 A d\tilde{\phi}=2 A_{\phi}d\tilde{\phi}= A_1dx_1+A_2 dx_2, ~~~ \delta_\Lambda A=2\delta_\Lambda A_{\phi}= \delta_{\Lambda} A_1+\delta_{\Lambda} A_2,
\label{phaseSpace}
 \end{eqnarray}
where the effective angular variable is $\tilde{\phi}=\frac{1}{2} \phi$.

By convenience we should choose the redefinition of generators (\ref{DiracChoice}) that was used during the explicit computation in section \ref{DiracBPrincipal}, namely
\begin{eqnarray}
\mathcal{L}^{(0)}_1&=& 2 \mathcal{W}^{(0)} +2 x_1 \partial_1 \mathcal{W}^{(0)},\nonumber\\
\mathcal{W}^{(0)}_1&=&-{\mathcal{L}^{(0)}}^2-\frac{1}{6}\partial^2_1 \mathcal{L}^{(0)}-x_1\frac{1}{6}\left(16 \mathcal{L}^{(0)}\partial_1
\mathcal{L}^{(0)}+\partial_1^3 \mathcal{L}^{(0)}\right). \nonumber
\end{eqnarray}
By the following redefinition of residual gauge parameters
\begin{eqnarray}
\epsilon^{(0)}_1&=&x_1 \left(\frac{8}{3}\mathcal{L}^{(0)}\partial_1 \eta^{(0)}+\frac{1}{6}\partial_1^3 \eta^{(0)}\right), \nonumber \\
\eta^{(0)}_1 &=&-2 x_1 \partial_1 \epsilon^{(0)}, \label{DiracChoice2}
\end{eqnarray}
we get rid of all terms in the residual gauge transformation $\delta_{\Lambda}A$ that break periodicity under $\phi\rightarrow \phi+2\pi$. 

With the choices above, the $V^2_{-1}$ and $V^3_{-2}$ components of $A$ become $\mathcal{L}^{(0)}+\frac{1}{2}\mu_3 t_0 \mathcal{L}^{(1)}+O(\mu_3^2)$ and $\mathcal{W}^{(0)}+\frac{1}{2}\mu_3 t_0 \mathcal{W}^{(1)}+O(\mu_3^2)$ respectively. The $(\mathcal{L}^{(1)},\mathcal{W}^{(1)})$ are determined by the equations of motion (\ref{x2equations}) to be
\begin{eqnarray}
\mathcal{L}^{(1)}&=&2 \partial_1 \mathcal{W}^{(0)},\nonumber\\
\mathcal{W}^{(1)}&=&-\frac{1}{6} \left(16 \mathcal{L}^{(0)}\partial_1
\mathcal{L}^{(0)}+\partial_1^3 \mathcal{L}^{(0)}\right).
\end{eqnarray}
Notice that explicit dependence in the Cauchy surface position $t_0$ remains in both $A$ and $\delta_{\Lambda}A$. The contribution of this explicit dependence in $t_0$ to the charge $Q$ is a total derivative whose integration vanishes upon imposing our periodic boundary conditions. The integrated charge, out of (\ref{boundarytNonLinear2}), for any $t_0$
\begin{equation}
Q(t_0)=\int^\pi_{0} d\tilde{\phi} \left(\epsilon^{(0)}\mathcal{L}^{(0)}-\eta^{(0)}\left(\mathcal{W}^{(0)}+\mu_3\left(\frac{1}{3}\partial^2_{1}\mathcal{L}^{(0)}
+\frac{1}{3}{\mathcal{L}^{(0)}}^2\right)\right)\right)+O(\mu_3^2),
\end{equation}
and the variations 
\begin{eqnarray}
\delta_\Lambda \mathcal{L}^{(0)}&=&\ldots + \mu_3\left(2\partial_1\mathcal{W}^{(0)}\epsilon^{(0)}+4\mathcal{W}^{(0)}\partial_1\epsilon^{(0)}+4 \mathcal{L}^{(0)}\partial_1\mathcal{L}^{(0)}\eta^{(0)}\right. \nonumber\\ && \left. ~~~~~~~~~~~+3{\mathcal{L}^{(0)}}^2\partial_1 \eta^{(0)}+3 \partial_1 \eta^{(0)}\partial_1^2\mathcal{L}^{(0)}+\frac{11}{2}\partial_1 \mathcal{L}^{(0)}\partial_1^2 \eta^{(0)}\right. \nonumber\\&&\left.+\frac{1}{3} \partial_1^3\mathcal{L}^{(0)}\eta^{(0)}+\frac{8}{3}\mathcal{L}^{(0)}\partial_1^3\eta^{(0)}+\frac{1}{6} \partial_1^5 \eta \right)+O(\mu_3^2),\label{deltalambda}\\
\delta_\Lambda \mathcal{W}^{(0)}&=&\ldots+ \mu_3\left(-\frac{8}{3}\mathcal{L}^{(0)}\partial_1\mathcal{L}^{(0)}\epsilon^{(0)}-\frac{13}{3}
{\mathcal{L}^{(0)}}^2\partial_1\epsilon^{(0)}-\frac{4}{3}\partial_1^2\mathcal{L}^{(0)}\partial_1\epsilon^{(0)}\right.\nonumber\\ &&\left.~~~~~~~~~~-\frac{25}{6} \partial_1\mathcal{L}^{(0)}\partial_1^2\epsilon^{(0)}-\frac{1}{6}\partial_1^3\mathcal{L}^{(0)}\epsilon^{(0)}-\frac{11}{3}\mathcal{L}^{(0)}\partial_1^3\epsilon^{(0)}-\frac{1}{3}\partial_1^5\epsilon^{(0)}\right. \nonumber\\&&~~~~~~~~~\left.+\frac{16}{3}\mathcal{W}^{(0)}\partial_1\mathcal{L}^{(0)}\eta^{(0)} +\frac{20}{3}\mathcal{L}^{(0)}\partial_1\mathcal{W}^{(0)}\eta^{(0)} +\frac{38}{3}\mathcal{L}^{(0)}\mathcal{W}^{(0)}\partial_1\eta^{(0)}\right. \nonumber\\&&\left. ~~~~~\frac{10}{3}\partial_1^2\mathcal{W}^{(0)}\partial_1\eta^{(0)}+\frac{11}{3}\partial_1\mathcal{W}^{(0)}\partial_1^2\eta^{(0)}+\frac{5}{3}\mathcal{W}^{(0)}\partial_1^3\eta^{(0)}+\partial_1^3\mathcal{W}^{(0)}\eta^{(0)}\right)  \nonumber\\ &&~~~~~~~~~~~~~~~~~~~~~~~~~~~~~~~~~~~~~~~~~~~~~~~~~~~~~~~~~~~~~~~~~~~~~~~~~+O(\mu_3^2), \nonumber\\
\delta_\Lambda \mathcal{L}^{(1)}&=&\left(\delta \mathcal{L}^{(1)}\right)|_{\delta \rightarrow \delta_\Lambda},\nonumber\\ \delta_\Lambda \mathcal{W}^{(1)}&=&\left(\delta \mathcal{W}^{(1)}\right)|_{\delta \rightarrow \delta_\Lambda}, \label{gaugeVARDirac}
\end{eqnarray}
determine, after long but straightforward computation, the fixed time $t_0$ Dirac bracket algebra (\ref{Dbras3}) by means of (\ref{rstra})\footnote{\ldots with the substitution $(x_1,\partial_1)\rightarrow(\frac{t_0}{2}+\tilde{\phi},\partial_{\tilde{\phi}})$ always implicitly intended.}. 

 The $\ldots$ in (\ref{deltalambda}) stand for the zeroeth order in $\mu_3$ contribution, which is given by the rhs of (\ref{gaugeVARIATIONS}) after substituting ($\mathcal{L},\mathcal{W},\epsilon,\eta$) by ($\mathcal{L}^{(0)},\mathcal{W}^{(0)},\epsilon^{(0)},\eta^{(0)}$) respectively. Remember that $\delta$ stands for arbitrary functional differential and so by $(\delta \ldots)|_{\delta\rightarrow\delta_\Lambda}$ we mean to take the functional differential of $\ldots$ in terms of $(\delta \mathcal{L}^{(0)},\delta \mathcal{W}^{(0)})$ and after substitute $\delta$ by $\delta_{\Lambda}$. 
 
 As we already said at the end of section \ref{DiracBPrincipal}, and stress again, the $\mu_3$ deformation of (\ref{Dbras3}) can not be absorbed by a field redefinition. In other words the fixed time Dirac bracket algebra (\ref{Dbras3}) is not isomorphic to $W_3$.
%

Notice that, and we must insist on this point, a different choice of field dependent redefinition of gauge parameter than (\ref{DiracChoice2}) would define a different (up to redefinition of generators) bracket algebra than the Dirac one (\ref{Dbras3}). This is, the new bracket algebra will not correspond to the $P$-phase space (up to coordinate redefinitions, of course) but to a different phase space. This is what the authors in \citep{Compere:2013gja} have done. We will review in our way their computations and will provide our interpretation of their results. 

\subsection{The change to $W_3$ of \citep{Compere:2013gja}} \label{redW3}

In this subsection we illustrate the issue mentioned in the previous paragraph. We will explicitly see that by using a field dependent redefinition of the gauge parameter different than \eqref{DiracChoice2} one alters the fixed time Dirac bracket algebra (of the original $P$-phase space) to an algebra isomorphic to $W_3$. This result could confuse the reader as one could naively think that with such a field dependent redefinition of the residual gauge parameter the initial phase space remains the same up to coordinate redefinitions. In fact, to complete this analysis, during the next subsection we will explicitly show that such a redefinition is nothing else but a non residual gauge transformation that maps the $P$-phase space (\ref{Dirichlet1}) onto the highest weight gauge phase space used in \citep{Campoleoni:2010zq
}.

The new choice
 \begin{eqnarray}
\mathcal{L}^{(0)}_1=\ldots+\mathcal{W}^{(0)},~~ \mathcal{W}^{(0)}_1=\ldots-\frac{5}{3} {\mathcal{L}^{(0)}}^2-\frac{7}{12}\partial_1^2\mathcal{L}^{(0)}, \nonumber\\
\epsilon^{(0)}_1=\ldots- \left(\frac{8}{3}\eta^{(0)} \mathcal{L}^{(0)}+\frac{1}{4}\partial_{1} ^2 \eta^{(0)}\right),~~
\eta^{(0)}_1=\ldots +\epsilon^{(0)}, \label{RedefPar}
\end{eqnarray}
instead of the previous ones (\ref{DiracChoice}) and (\ref{DiracChoice2}), with the $\ldots$ denoting the rhs of the respective (\ref{DiracChoice}) and (\ref{DiracChoice2}) expressions,  defines the integrated charge
\begin{equation}
Q(t_0)=\int_{0}^\pi d \tilde{\phi}  \left(\epsilon^{(0)} \mathcal{L}^{(0)}-\eta^{(0)}\mathcal{W}^{(0)} \right)+O(\mu_3^2), \label{chargeW3}
\end{equation}
with variations ($\delta_{\Lambda}\mathcal{L}^{(0)},\delta_\Lambda \mathcal{W}^{(0)}$) given precisely as in (\ref{gaugeVARIATIONS}) with ($\mathcal{L},\mathcal{W},\epsilon,\eta$) substituted by the initial conditions ($\mathcal{L}^{(0)},\mathcal{W}^{(0)},\epsilon^{(0)},\eta^{(0)}$). 

The variations ($\delta_{\Lambda}\mathcal{L}^{(1)},\delta_\Lambda\mathcal{W}^{(1)}$) are given in terms of ($\delta_{\Lambda}\mathcal{L}^{(0)},\delta_\Lambda\mathcal{W}^{(0)}$),  as presented in the last two lines in (\ref{gaugeVARDirac}). Thence from (\ref{rstra}) one derives (\ref{ASA2}) which is $W_3$. As already stated this Poisson structure is not equivalent to the Dirac structure (\ref{Dbras3}) mentioned before.
The technical reason being the presence of the field dependent redefinition of gauge parameters (\ref{RedefPar}) that is not equivalent to a redefinition of ($\mathcal{L}^{(0)},\mathcal{W}^{(0)}$). As we will show this procedure is somehow violating the Dirichlet boundary conditions (\ref{Dirichlet1}).

But before going on let us write down the expression for the original $(V^{2}_{-1},V^{3}_{-2})$ components of the projection $A_1$ of $A$ and the corresponding residual gauge parameters, ($\mathcal{L},\mathcal{W},\epsilon,\eta$), in terms of the ($\mathcal{L}^{(0)},\mathcal{W}^{(0)},\epsilon^{(0)},\eta^{(0)}$) for the choice (\ref{RedefPar})

\begin{eqnarray}
\mathcal{L}&=&\mathcal{L}^{(0)}+3 \mu_3 \mathcal{W}^{(0)}+\mu_3 t_0 \partial_1 \mathcal{W}^{(0)}+O(\mu_3^2),\nonumber\\ \mathcal{W}&=&\mathcal{W}^{(0)}-\mu_3\left(\frac{8}{3}{\mathcal{L}^{(0)}}^2+\frac{3}{4}\partial_{x_1}^2 \mathcal{L}^{(0)}\right)-\frac{1}{12}\mu_3 t_0\left(16 \mathcal{L}^{(0)}\partial_1
\mathcal{L}^{(0)}+\partial_1^3 \mathcal{L}^{(0)}\right)+O(\mu_3^2), \nonumber\\
\epsilon&=&\epsilon^{(0)}-\mu_3\left(\frac{8}{3}\eta^{(0)} \mathcal{L}^{(0)}+ \frac{1}{4}\partial_{x_1}^2\eta^{(0)}\right)+\frac{1}{12}\mu_3 t_0 \left(16 \mathcal{L}^{(0)}\partial_1 \eta^{(0)}+\partial_1^3 \eta^{(0)}\right) +O(\mu_3^2), \nonumber\\\eta&=&\eta^{(0)}+\mu_3 \epsilon^{(0)}-\mu_3 t_0 \partial_1 \epsilon^{(0)}+O(\mu_3^2). \label{SongTrans}
\end{eqnarray}
\\
The $(V^{2}_{-1}, V^{3}_{-2})$ components of $A$ are recovered by dropping the terms linear in $\mu_3$ without $t_0$ dependence in the first two lines in (\ref{SongTrans}).\\

\subsubsection{The change to $W_3$ of \citep{Compere:2013gja} as a non residual transformation to the highest weight gauge} \label{RTW3}
As promised, we will show that the process that follows the choice (\ref{RedefPar}) in defining a $W_3$ algebra, is equivalent to the process of performing a non residual gauge transformation\footnote{This argument has been already presented by the authors in \citep{Bunster:2014mua}. Here we provide this instance from our own perspective.} that maps the $P$-phase space (\ref{Dirichlet1}) to the highest weight gauge phase space used in \citep{Campoleoni:2010zq}. In other words it is equivalent to perform a gauge transformation that changes the original boundary conditions and thence the new $W_3$ bracket algebra, corresponds to a different phase space, not to (\ref{Dirichlet1}). 

Firstly, let us discuss some facts that will be useful in reaching our purpose. Let $A$ be the space of flat connections with residual gauge transformation condition $\delta A=D_A\Lambda_A$. 

Let $g$ be an arbitrary field dependent gauge group element which is not a residual transformation of $A$. By performing the similarity transformation by $g$ on both sides of $(\delta A)= D_A\Lambda_A $ we get
\begin{eqnarray}
g \delta A g^{-1}&=&\delta A_g-D_{A_g}(g\delta g^{-1}),\nonumber \\
g D_A (\Lambda_A)g^{-1}&=&D_{A_g}(g \Lambda_A g^{-1}),\label{transRes}
\end{eqnarray}
where $A_g\equiv g A g^{-1}+g \partial g^{-1}$. From (\ref{transRes}) we read out the transformation law for the residual gauge parameter $\Lambda$
\begin{eqnarray}
\Lambda_{A_g}=g \Lambda_{A}g^{-1}+g\delta g^{-1}, \label{Map}
\end{eqnarray}
where at this point, we are free to substitute the arbitrary differential $\delta$ by $\delta_{\Lambda_A}$, the initial residual gauge transformation. 

Now we notice that equations (\ref{x2equations}) and (\ref{x2equationsPar}) are integrable at any order in $\mu_3$ as it follows from gauge invariance \citep{Ferlaino:2013vga, Compere:2013gja}.  One way to solve them is to express the solution in terms of a gauge group element $g=g(\mathcal{\tilde{L}},\mathcal{\tilde{W}}, \mu_3 x_2)$  that takes the highest weight connection
  \begin{eqnarray}
    \tilde{A}_1= V^{2}_1+\tilde{\mathcal{L}} V^{2}_{-1}+\tilde{\mathcal{W}} V^{3}_{-2}, ~~ \tilde{A}_2=0, \label{InitialCond}
    \end{eqnarray}
 to (\ref{Dirichlet1}), via the gauge transformation law $\tilde{A} \rightarrow \tilde{A}_g \equiv A$. The element $g$ that transforms (\ref{InitialCond}) into (\ref{Dirichlet1}) is generated at the first order in $\mu_3$ and linear order in  the algebra element by:
 \begin{eqnarray}
 \Lambda_g&=&\Lambda(\tilde{\epsilon}_g,\tilde{\eta}_g)-x_2 A_2+O(\mu_3^2) \nonumber\\
 &=&\Lambda(\tilde{\epsilon}_g,\tilde{\eta}_g)+ \Lambda(0,-\mu_3 x_2)+O(\mu_3^2),
 \label{gaugeParameterTheisen}
 \end{eqnarray}
 with $\Lambda$, as a function of ($\tilde{\epsilon},\tilde{\eta}$), given by (\ref{gaugeParameter}) with background fields ($\tilde{\mathcal{L}},\tilde{\mathcal{W}}$) instead of ($\mathcal{L},\mathcal{W}$). From the second line in (\ref{gaugeParameterTheisen}) it follows that $\Lambda_g$ generates transformations of the kind (\ref{gaugeVARIATIONS}) on the $(\tilde{\mathcal{L}},\tilde{\mathcal{W}})$ and relate them with the new parameters $(\mathcal{L},\mathcal{W})$ by
 \begin{eqnarray}
 \mathcal{L}=\tilde{\mathcal{L}}-2\mu_3 x_2 \partial_1\tilde{\mathcal{W}}+O(\mu_3^2),~ \mathcal{W}=\tilde{\mathcal{W}}+\mu_3 x_2\left(\frac{8}{3}\tilde{\mathcal{L}}^2+\frac{1}{6}\partial_1^2\tilde{\mathcal{L}}\right)+O(\mu_3^2),\label{Identification}
 \end{eqnarray}
where we have hidden the arbitrariness $\Lambda(\tilde{\epsilon}_g,\tilde{\eta}_g)$ in (\ref{gaugeParameterTheisen}), inside of the $(\tilde{\mathcal{L}},\tilde{\mathcal{W}})$. From the $x_2$ flow equations (\ref{x2equations}) and (\ref{Identification}) one is able to identify the parameters $(\tilde{\mathcal{L}},\tilde{\mathcal{W}})$ with the initial conditions
\begin{equation}
\tilde{\mathcal{L}}\equiv\mathcal{L}^{(0)}+\mu_3 \mathcal{L}^{(0)}_1+O(\mu_3^2),~ \tilde{\mathcal{W}}\equiv\mathcal{W}^{(0)}+\mu_3 \mathcal{W}^{(0)}_1+O(\mu_3^2). \label{Constante}
\end{equation}
The gauge transformation induced by $g$ is then identified with the hamiltonian evolution along $x_2$ that recovers $(\mathcal{L},\mathcal{W})$ out of the initial conditions (\ref{Constante}).

 Now we can apply (\ref{Map}) to this specific case
   \begin{eqnarray}
    \Lambda&=&g\tilde{\Lambda}g^{-1}+ g \delta g^{-1}\nonumber\\&=&\tilde{\Lambda}+x_2\left(\delta A_2-[A_2,\Lambda]\right)+O(\mu_3^2)=\tilde{\Lambda}+ x_2 \partial_2\Lambda|_{x_2=0}+O(\mu_3^2) \nonumber\\&=&\tilde{\Lambda}+ x_2\left(-\mu_3\left(\frac{8}{3}\tilde{\mathcal{L}}\partial_1\tilde{\eta}+\frac{1}{6}\partial_1^3\tilde{\eta}\right)V^{2}_1+2 \mu_3 \partial_1 \tilde{\epsilon} V^3_2+\ldots\right)+O(\mu_3^2).\nonumber\\ \label{Identification2}
    \end{eqnarray}
Where by $\delta$ we mean the analog of the variations  (\ref{gaugeVARIATIONS}), and again we have hidden the arbitrariness $\Lambda(\tilde{\epsilon}_g,\tilde{\eta}_g)$ inside the parameters $\tilde{\Lambda}\equiv\Lambda(\tilde{\epsilon},\tilde{\eta})$. The last line in (\ref{Identification2}), together with the $x_2$ flow equations (\ref{x2equationsPar}), allows us to identify the parameters $(\tilde{\epsilon},\tilde{\eta})$ with the initial conditions $(\epsilon^{(0)}+\mu_3\epsilon^{(0)}_1+O(\mu_3^2),\eta^{(0)}+\mu_3\eta^{(0)}_1+O(\mu_3^2))$. For later reference
 \begin{equation}
 \tilde{\epsilon}\equiv\epsilon^{(0)}+\mu_3 \epsilon^{(0)}_1+O(\mu_3^2), ~ \tilde{\eta}\equiv\eta^{(0)}+\mu_3 \eta^{(0)}_1+O(\mu_3^2).\label{Constante2}
 \end{equation}
After imposing (\ref{RedefPar}), the explicit form of $\Lambda$  (\ref{gaugeParameter}), (\ref{Constante}), (\ref{Constante2}) on (\ref{Identification}) and (\ref{Identification2}), one finds the same expressions (\ref{RedefPar}) gotten from the previous procedure for $(\mathcal{L}^{(0)}_1,\mathcal{W}^{(0)}_1,\epsilon^{(0)}_1$, $\eta^{(0)}_1)$. 

We have thence proven that the process that follows the choice (\ref{RedefPar}) in defining a $W_3$ algebra, is equivalent to the process of performing the non residual gauge transformation (\ref{gaugeParameterTheisen}) that maps the $P$-phase space (\ref{Dirichlet1}) to the highest weight gauge phase space (\ref{InitialCond}) used in \citep{Campoleoni:2010zq}.


Finally, let us provide a different perspective to understand the significance of the choice of $\mu_3$ dependence, $(\mathcal{L}^{(0)}_1,\mathcal{W}^{(0)}_1,\epsilon^{(0)}_1,\eta^{(0)}_1)$, in the integration constants $(\tilde{\mathcal{L}},\tilde{\mathcal{W}}, \tilde{\epsilon}, \tilde{\eta})$. From (\ref{Map}) it follows that the differential of charge $\delta Q\equiv\int^{\pi}_0 d\tilde{\phi}~ tr(\tilde{\Lambda}\delta A)$ is not invariant under a generic gauge transformation. In particular, the differential of charge for (\ref{InitialCond}) previous to the gauge transformation $g$ encoding the $x_2$ evolution, is:
\begin{equation}
\delta Q(\tilde{\epsilon},\tilde{\eta})\equiv\int^{\pi}_0 d\tilde{\phi}~ tr(\tilde{\Lambda}\delta\tilde{A}_1)=\int^{\pi}_0 d\tilde{\phi}\left( \tilde{\epsilon} \delta \tilde{\mathcal{L}}-\tilde{\eta} \delta \tilde{\mathcal{W}}\right),
\end{equation}
and picks up an extra $\mu_3$ dependence after a generic $\mu_3$ dependent non residual gauge transformation is performed. The choice (\ref{RedefPar}) is the one that cancels, up to trivial integrations of total derivatives, the extra $\mu_3$ dependence contribution to the final differential of charge. The final result for the transformed charge,  after functional integration is performed, coincides with (\ref{chargeW3}). This result is a consequence of the fact that the transformation $g$ to the highest weight gauge is equivalent to perform the field dependent redefinition (\ref{RedefPar}).

 Notice that in consequence, the non residual gauge transformation $g$ takes to a phase space (\ref{InitialCond}) different than the $P$-phase space (\ref{Dirichlet1}). 
As this non residual gauge transformation $g$ is equivalent to the choice (\ref{RedefPar}) we have thence proven that the field dependent redefinition of residual gauge parameter (\ref{RedefPar}) does not preserve the form of the $P$-phase space. So, the $W_3$ algebra obtained after performing (\ref{RedefPar}) does not act onto the $P$-phase space  (\ref{Dirichlet1}). In consequence, the existence of the change (\ref{RedefPar}) to a $W_3$ algebra \citep{Compere:2013gja}, is not in contradiction at all, with the fact that the fixed time Dirac bracket algebra, aka fixed time asymptotic symmetry algebra, computed for the $P$-phase space (\ref{Dirichlet1}) is not isomorphic to $W_3$. 

\subsection{Explicit computation of Dirac bracket algebra in $D$-phase space} \label{DiracBDiagonal}
In this section we try to identify a $W^{(2)}_3$ Dirac bracket structure of another phase space that contains black holes \citep{Bunster:2014mua}.

Firstly, we review how to embed the $P$-phase space (\ref{Dirichlet1}) into a larger phase space
. We call it $D$-phase space after the fact we use the diagonal ($D$) embedding classification of generators to describe it \footnote{We use these $P,~D$ prefixes to stress the difference between both phase spaces.}. Finally we compute the corresponding fixed time Dirac bracket algebra and show that it is isomorphic to $W^{(2)}_3$. 

First we redefine our generators as
\begin{eqnarray}
J_0=\frac{1}{2} V^{2}_0, \  J_{\pm}=\pm \frac{1}{2} V^{3}_{\pm2} , \  \Phi_{0}=V^{3}_0,~~~~~~~~~~~~~\nonumber\\
 ~~~~G^{(\pm)}_{ \frac{1}{2}}= \frac{1}{\sqrt{8}}\left( V^{2}_1\mp 2 V^{3}_1\right), ~~ G^{(\pm)}_{ -\frac{1}{2}}= -\frac{1}{\sqrt{8}}\left( V^{2}_{-1}\pm 2 V^{3}_{-1}\right), \label{basisW32}
\end{eqnarray}
with the non trivial commutation relations being:
\begin{eqnarray}
[J_i,J_j]=(i-j)J_{i+j}, ~~[J_i,\Phi_0]=0, ~~ [J_{i},G^{(a)}_{m}]=(\frac{i}{2}-m) G^{(a)}_{i+m},\nonumber \\
~[ \Phi_0,G^{(a)}_m]=a G^{(a)}_m, ~~[G^{(+)}_m,G^{(-)}_n]=J_{m+n}-\frac{3}{2}(m-n)\Phi_0,~~~~~~
\end{eqnarray}
with $i=-1,0,1$, $m=-\frac{1}{2},\frac{1}{2}$ and $a=\pm$. The $J$'s denoting the $sl(2,\mathbb{R})$ generators in the diagonal embedding. After the shift $\rho\rightarrow \rho-\frac{1}{2}\log(\mu_3)$, the space of flat connections (\ref{Dirichlet1}) can be embedded into
\begin{eqnarray}
A_1&=&\nu_3 \left(\sqrt{2}\left(G^{(+)}_{\frac{1}{2}}+G^{(-)}_{\frac{1}{2}}\right)-\frac{1}{\sqrt{2}} \left(\mathcal{G}^++\mathcal{G}^-\right)J_{-}-\sqrt{3}\mathcal{J}\left(G^{(+)}_{-\frac{1}{2}}+G^{(-)}_{-\frac{1}{2}}\right)\right),\nonumber\\
A_2&=&2 J_{+}+2\mathcal{G}^+G^{(+)}_{-\frac{1}{2}}+2\mathcal{G}^-G^{(-)}_{-\frac{1}{2}}+ \sqrt{6}\mathcal{J} \Phi_0+2 \mathcal{T}^\prime J_-, \label{family2}
\end{eqnarray}
where $\nu_3\equiv\mu_3^{-\frac{1}{2}}$ and
\begin{eqnarray}
\mathcal{G}^{+}&=&\frac{\sqrt{2}}{6}\mu_3^{\frac{3}{2}}\left(\partial_1 \mathcal{L}+6 \mathcal{W}\right), ~\mathcal{G}^-=-\frac{\sqrt{2}}{6}\mu_3^{\frac{3}{2}}\left(\partial_1 \mathcal{L}-6 \mathcal{W}\right),\nonumber\\ ~ \mathcal{J}&=& \sqrt{\frac{2}{3}} \mu_3\mathcal{L}, ~ \mathcal{T}^\prime= -\frac{1}{6}\mu_3^2(\partial_1^2 \mathcal{L}+6 \mathcal{L}^2).~~~~~~~~~~ \label{changeW32}
\end{eqnarray}

To obtain the previous phase space (\ref{Dirichlet1}) out of (\ref{family2}), one must impose restrictions on the latter. This is, relations (\ref{changeW32}) imply the constraints
\begin{eqnarray}
\mathcal{G}^+-\mathcal{G}^--\frac{1}{\sqrt{3}~\nu_3}\partial_1 \mathcal{J}=0, ~ \mathcal{T}^\prime+\frac{1}{2\sqrt{6}~ \nu_3^2}\left(\partial_1^2\mathcal{J}+\nu_3^2\sqrt{\frac{3}{2}}\mathcal{J}^2\right)=0,\label{Intertwine}
\end{eqnarray}
which are not compatible with the equations of motion
\begin{eqnarray}
\partial_1 \mathcal{G}^\pm&=&\mp \frac{\nu_3}{2\sqrt{2}} \left(6 \mathcal{J}^2\pm\sqrt{6}\partial_2 \mathcal{J}+4 \mathcal{T}^\prime\right), ~ \partial_1\mathcal{J}=\sqrt{3} \nu_3\left(\mathcal{G}^+-\mathcal{G}^{-}\right),\nonumber\\ \partial_1\mathcal{T}^\prime&=&-\nu_3\left(\sqrt{3}\left(\mathcal{G}^- -\mathcal{G}^+\right)\mathcal{J}+\frac{1}{2\sqrt{2}}\left(\partial_2 \mathcal{G}^-+\partial_2\mathcal{G}^+\right) \right),\label{EoMW32}
\end{eqnarray}
and hence define second class constraints on the corresponding phase space of solutions. We will not impose them, in fact they are non perturbative in $\nu_3$. As already mentioned, we will denote the phase space (\ref{family2}) with the prefix $D$.

The gauge parameter of residual gauge transformations for (\ref{family2})
\begin{eqnarray}
\Lambda&=&2\Lambda_{J_{+}}J_{+}+ 2 \Lambda_{G^{+}_{\frac{1}{2}}} G^{+}_{\frac{1}{2}}+ 2\Lambda_{G^{-}_{\frac{1}{2}}} G^{-}_{\frac{1}{2}}+\sqrt{6}\Lambda_{\Phi_0} \Phi_0~~~~~~~~~~~~~~~~~~~~~~\nonumber \\&~&+
\left(-\frac{1}{2}\partial_2 \Lambda_{J_+}\right)J_{0}+\left(-\mathcal{G}^{+} \Lambda_{G^{(-)}_{\frac{1}{2}}}-\mathcal{G}^{-} \Lambda_{G^{(+)}_{\frac{1}{2}}}+2\mathcal{T}^\prime\Lambda_{J_+}+\frac{1}{4}\partial_2^2 \Lambda_{J_+}\right)J_{-}\nonumber\\&~&+\left(-\sqrt{6}\mathcal{J}\Lambda_{G^{+}_{\frac{1}{2}}}+2\mathcal{G}^{(+)} \Lambda_{J_{+}}-\partial_2 \Lambda_{G^{(+)}_{\frac{1}{2}}}\right)G^{+}_{-\frac{1}{2}}\nonumber\\&~&+\left(-\sqrt{6}\mathcal{J}\Lambda_{G^{{-}}_{\frac{1}{2}}}+2\mathcal{G}^{(-)} \Lambda_{J_+}+\partial_2 \Lambda_{G^{(-)}_{\frac{1}{2}}}\right)G^{-}_{-\frac{1}{2}},\nonumber \\
\end{eqnarray}
defines the variations
\begin{eqnarray}
\delta_{\Lambda_{J_+}} \mathcal{T}^\prime&=& \Lambda_{J_+} \partial_2 \mathcal{T}^\prime+2\partial_2 \Lambda_{J_+} \mathcal{T}^\prime+\frac{1}{8}\partial_2^3 \Lambda_{J_+}, \nonumber \\
\delta_{\Lambda_{\Phi_0}} \mathcal{J} &=&  \partial_2 \Lambda_{\Phi_0}, ~~ \delta_{\Lambda_{{G^{(+)}_{\frac{1}{2}}}}} \mathcal{J}=-\sqrt{6}\Lambda_{G^{(+)}_{\frac{1}{2}}} \mathcal{G}^-, ~~ \delta_{\Lambda_{G^{(-)}_{\frac{1}{2}}}} \mathcal{J}=\sqrt{6}\Lambda_{G^{(-)}_{\frac{1}{2}}} \mathcal{G}^+ ,\nonumber\\~ \delta_{\Lambda_{J_+}} \mathcal{G}^{(\pm)}&=&\partial_2\Lambda_{J_{+}} G+\frac{3}{2}\Lambda_{J_+}\partial_2 \mathcal{G}^{(\pm)}\pm\sqrt{6}\Lambda_{J_+}\mathcal{J}\mathcal{G}^{\pm},\nonumber \\
\delta_{\Lambda_{G^{+}_{\frac{1}{2}}}} \mathcal{G}^-&=&\left(2\mathcal{T}^\prime+3 \mathcal{J}^2-\sqrt{\frac{3}{2}}\partial_2 \mathcal{J}\right)\Lambda_{G^{+}_{\frac{1}{2}}}-\sqrt{6} \mathcal{J}\partial_2 \Lambda_{G^+_{\frac{1}{2}}}+\frac{1}{2}\partial_2 \Lambda_{G^+_{\frac{1}{2}}},\label{gaugeVARIATIONS5}
\end{eqnarray}
and the following differential of charge in the case of $x_1$ evolution
\begin{equation}
\delta Q=\int dx_2 tr\left(\Lambda \delta A_2\right)=\int dx_2 \left(\Lambda_{J_+} d\mathcal{T}-\Lambda_{\Phi_0}d\mathcal{J}-\Lambda_{G^{(-)}_{\frac{1}{2}}}d\mathcal{G}^+-\Lambda_{G^{(+)}_{\frac{1}{2}}}d\mathcal{G}^-\right). \label{chargeW32}
\end{equation}
We could now repeat the method of variation of generators done for the case of the principal embedding to this case, but instead we choose to work out the explicit computation of Dirac bracket algebra.

For the sake of brevity we will work at $t_0=0$, but the conclusion of this computation remains unchanged at any other fixed time slice. The difference being that the charges will carry an explicit $t_0$ dependence as in the previous case. At $t_0=0$ the Cauchy data at first order in $\nu_3$ can be written in the form
\begin{eqnarray}
A &=&2A_\phi d \tilde{\phi}=\left(A_{x_1}d x_1+A_{x_2}dx_2\right)\nonumber\\&=& \left(2 J_{+}+\sqrt{2} \nu_3\left(G^{(+)}_{-\frac{1}{2}}+G^{(-)}_{-\frac{1}{2}}\right)+2\tilde{\mathcal{G}}^{+(0)}G^{(+)}_{-\frac{1}{2}}+2\tilde{\mathcal{G}}^{-(0)}G^{(-)}_{-\frac{1}{2}}\right.\nonumber\\&&~~~~~~~~~~~~~~~~~~~~~~~~~~~\left.+ \sqrt{6}\mathcal{J}^{(0)} \Phi_0+2 \tilde{\mathcal{T}}^{\prime(0)}J_-\right) d\tilde{\phi}+O(\nu_3^2), \label{phaseW32}
\end{eqnarray}
by a choice of integration constants. Where
\begin{eqnarray}
\tilde{\mathcal{G}}^{\pm (0)}=\mathcal{G}^{\pm (0)}-\frac{\sqrt{3}}{2}\nu_3\mathcal{J}^{(0)}, ~\tilde{\mathcal{T}}^{\prime (0)}=\mathcal{T}^{\prime(0)}-\frac{1}{2\sqrt{2}}\nu_3\left(\mathcal{G}^{+ (0)}+\mathcal{G}^{- (0)}\right).
\end{eqnarray}
Again, we remind that by the super index $(0)$ we refer to the initial conditions of the system of $x_1$ evolution equations (\ref{EoMW32}). Some comments on notation are in order. Let the components of $A$ in the $W^{(2)}_3$ basis (\ref{basisW32}), be denoted again by $A_a$ with $a=1,\ldots,8$ and the ordering corresponding to
\begin{equation}
\left(J_0,J_+,J_-,\Phi_0,G^{(+)}_{-\frac{1}{2}},G^{(-)}_{-\frac{1}{2}},G^{(-)}_{-\frac{1}{2}},G^{(+)}_{-\frac{1}{2}}\right). \label{basisW32N}
\end{equation}
At this point, we impose the four second class constraints
\begin{equation}
C^i=\left(A_1,A_2-2,A_7-\sqrt{2} \nu_3 , A_8-\sqrt{2} \nu_3\right),\end{equation}
on the phase space (\ref{phaseW32}) endowed with the algebra (\ref{KMAlgebra}) written in the basis (\ref{basisW32N}). Notice that we shall not impose the second class constraints coming from (\ref{Intertwine}). As already mentioned they are non perturbative in $\nu_3$. 

Next, is straightforward to compute the Dirac bracket (\ref{DBras}). For completeness we write down the non vanishing elements of $M_{ij}$ in this case
\begin{eqnarray}
M_{11}=\frac{1}{8}\partial_{x_2} \delta_{x_2y_2}, ~ M_{12}=-M_{21}=-\frac{1}{2\sqrt{2}}\delta_{x_2y_2}, ~ M_{34}=-M_{43}=\frac{1}{2}\delta_{x_2y_2},\nonumber\\ ~ M_{13}=-M_{31}=M_{41}=-M_{14}= \frac{\nu_3}{4 \sqrt{2}} \delta_{x_2y_2},  ~~~~~~~~~~~~~~~
\end{eqnarray}
from where we can check explicitly by using (\ref{DBras}) that $\{C^i,\ldots\}_D=0$. 

The algebra in the reduced phase space will depend on $\nu_3$ explicitly, but after implementing the change
\begin{eqnarray}
\mathcal{G}_{\nu_3}^{\pm (0)}=\tilde{\mathcal{G}}^{\pm (0)}-\frac{\sqrt{3}}{2} \nu_3\mathcal{J}^{(0)}, ~\mathcal{T}_{\nu_3}^{\prime}=\tilde{\mathcal{T}}^{\prime}-\frac{1}{\sqrt{2}}\nu_3( \tilde{\mathcal{G}}^{+ (0)}+\tilde{\mathcal{G}}^{-(0)}),
\end{eqnarray}
we obtain the undeformed $W^{(2)}_3$ algebra:
\begin{eqnarray}
\{ \mathcal{T}_{\nu_3}^{\prime (0)}(y_2),\mathcal{T}_{\nu_3}^{\prime(0)}(x_2)\}_D&=&\mathcal{T}_{\nu_3}^{\prime (0)} \delta_{x_2y_2}+2 \partial_{x_2} \mathcal{T}_{\nu_3}^{\prime(0)} \delta_{x_2y_2}+\frac{1}{8}\partial_{x_2} \delta_{x_2y_2},\nonumber\\
\{\mathcal{J}^{(0)}_{\nu_3}(y_2),\mathcal{J}^{(0)}_{\nu_3}(x_2)\}_D&=&\delta_{x_2y_2},\nonumber\\ \{\mathcal{J}^{(0)}_{\nu_3}(y_2),\mathcal{G}^{\pm (0)}_{\nu_3}(x_2)\}_D&=&\pm\sqrt{6} \mathcal{G}^{\pm (0)}_{\nu_3} \delta_{x_2y_2},\nonumber \\
\{\mathcal{T}_{\nu_3}^{\prime(0)}(y_2), \mathcal{G}^{\pm (0)}_{\nu_3}(x_2)\}_D&=& \partial_{x_2} \mathcal{G}^{\pm (0)}_{\nu_3}\delta_{x_2 y_2}+\frac{3}{2}\mathcal{G}^{\pm (0)}_{\nu_3}\partial_{x_2}\delta_{x_2 y_2}\pm \sqrt{6} \mathcal{J}^{(0)}_{\nu_3} \mathcal{G}^{\pm (0)}_{\nu_3}\delta_{x_2 y_2}, \nonumber\\
\{\mathcal{G}^{+ (0)}_{\nu_3}(y_2),\mathcal{G}^{- (0)}_{\nu_3}(x_2)\}_D&=&-\left(2 \mathcal{T}^{\prime 0}_{\nu_3}+3 {\mathcal{J}^{(0)}_{\nu_3}}^2-\sqrt{\frac{3}{2}}\partial_{x_2}\mathcal{J}^{(0)}_{\nu_3}\right) \delta_{x_2 y_2}\nonumber\\&~&+\sqrt{6}\mathcal{J}^{(0)}_{\nu_3}\partial_{x_2}\delta_{x_2y_2} -\partial_{x_2}^2 \delta_{x_2y_2},\nonumber \\
\end{eqnarray}
that agrees precisely with the signature of charges in (\ref{chargeW32}) and the transformation laws (\ref{gaugeVARIATIONS5}). The most canonical form can be achieved by the usual redefinition of energy momentum tensor $\mathcal{T}_{\nu_3}^{\prime (0)}\rightarrow \mathcal{T}_{\nu_3}^{\prime (0)}+\frac{1}{2} {\mathcal{J}^{(0)}_{\nu_3}}^2$ that makes $G^{\pm (0)}_{\nu_3}$ and $\mathcal{J}^{(0)}_{\nu_3}$ primaries of weight $\frac{3}{2}$ and $1$ respectively.
It is then proven that the fixed time asymptotic symmetry algebra of the space of solutions (\ref{family2}) is $W^{(2)}_3$ at first order in the parameter $\nu_3$ \footnote{However, this should be the case at any order in $\nu_3$. As suggested by the non perturbative analysis reported in appendix B.2 of \citep{Bunster:2014mua}. Notice that to compute explicitly Dirac brackets we were forced to the use of perturbation theory. For an alternative non perturbative analysis, the reader can refer to \citep{Bunster:2014mua}. }.

Notice that \eqref{family2} does contain the $(\mu_3, \bar{\mu}_3)$ black hole solutions \citep{Gutperle:2011kf} (of course, after performing the shift $\rho\rightarrow \rho-\frac{1}{2}\log(\mu_3)$ on them), as zero modes. Thence, both families \eqref{Dirichlet1} and \eqref{family2} can be used to define the charges of these black holes. However, the two possibilities are not equivalent as  we have already shown that \eqref{family2} is larger than \eqref{Dirichlet1} and thence the corresponding algebras are not isomorphic. The family \eqref{family2} is the preferred one, as for \eqref{Dirichlet1} it is impossible to define a basis of primary operators
\footnote{One can define a quasi-primary field of dimension 2, as a Virasoro subalgebra can be identified in \eqref{Dbras3}, but the remaining generator can not be redefined in order to form a primary with respect to the Virasoro one.}.

 We make a last comment before concluding. Notice that should we have worked with the following coordinates 
\begin{equation}
x_1=\frac{t+\phi}{2}, ~ x_2=\frac{\phi}{2}, \label{coordinates}
\end{equation}
all previously done remains valid, up to dependence on $t_0$. This dependence only affects implicitly the $W_3^{(2)}$ algebra through field redefinitions. The $hs(\lambda)$ ans\"atze introduced in \citep{Cabo-Bizet:2014wpa}, belong to (\ref{family2}) under (\ref{coordinates}) for the truncation to $sl(3,\mathbb{R})$ via the limit $\lambda=3$\footnote{However one should keep in mind the extra shift in the coordinate $\rho\rightarrow \rho-\frac{1}{2}\log(\mu_3)$
.}. Thenceforth, in this case, the corresponding charges are not of higher spin character.  

In our study we did not attempt to meddle with the issue of asymptotic symmetry algebras coming from generalised boundary conditions in the context of $hs(\lambda)$. We hope to report on that point in the near future.  

\section{Final remarks}
We started by analysing the Dirac bracket algebra on the phase space of $sl(3,\mathbb{R})$ CS in principal embedding (\ref{Dirichlet1}) after imposing the set of 6 constraints (\ref{Constmu3}) onto the corresponding Kac Moody algebra (\ref{KMAlgebra}) with $x_1=\frac{t+\phi}{2}$ and $x_2=\frac{-t+\phi}{2}$. 
Apart from the explicit computation, we used the method of variation of generators to cross check our result. The fixed time Dirac bracket algebra is not isomorphic to $W_3$. 


To complete our study, and try to elucidate the apparent contradiction, we have shown that the $W_3$ algebra that one can arrive to after a given field dependent redefinition of the smearing gauge parameter, as shown in \citep{Compere:2013gja} and here verified, does not act onto the original $P$-phase space (\ref{Dirichlet1}), but onto a phase space defined by a highest weight choice \citep{Campoleoni:2010zq,Campoleoni:2011hg,Henneaux:2010xg}.
 

Finally, we computed the fixed time Dirac bracket algebra in phase space (\ref{family2}), containing black holes, and as expected it turned out to be isomorphic to $W^{(2)}_3$ \citep{Compere:2013gja,Bunster:2014mua}. 
  
 It would be necessary to address similar questions for a generic value of the deformation parameter $\lambda$. For that, analysis in perturbations of the generalised boundary conditions in the corresponding phase spaces, like the expansion in $(\mu,\bar{\mu})$ in the $P$-phase space, or $(\nu,\bar{\nu})$ in the $D$-phase space of the $\lambda=3$ truncation here reviewed, could result helpful. 
Nevertheless we believe that an alternative and more general path to follow can be developed.\\ \\ 

\paragraph{Acknowledgments} 
We would like to thank Professors E. Gava and K. S. Narain for initial collaboration and revision of the manuscript. \\ \\

\appendix
\section{Conventions} \label{app:Conv}
The construction of the $hs(\lambda)$ algebra can be seen for example in \citep{Gaberdiel:2012uj}. The algebra is spanned by the set of generators $V^{s}_{t}$ with $s=2,\ldots, \infty$ and $1-s \leq t\leq s-1$. The element $V^1_0$ denotes the identity operator. To define the algebra we use the $\star$-product representation constructed in \citep{Pope:1990kc}:
\begin{equation}
\label{eq:lonestar}
V^{s}_{m}\star V^{t}_{n}=\frac{1}{2}\mathlarger{\mathlarger{\mathlarger{\sum}}}^{s+t-Max[\left|m+n\right|,\left|s-t\right|]-1}_{i=1,2,3,\dots} g^{s t}_{i}(m,n;\lambda) V^{s+n-i}_{m+n}
\end{equation}
With the constants:
\begin{equation}
\label{eq:strcons}
g^{s t}_{i}(m,n;\lambda)\equiv \frac{q^{i-2}}{2(i-1)!} 
\,_4F_3 \left[
\begin{array}{llll} 
\frac{1}{2}+\lambda & \quad \frac{1}{2}-\lambda \quad \frac{2-i}{2}\quad \frac{1-i}{2}\\
\frac{3}{2}-s & \quad \frac{3}{2}-t \quad \frac{1}{2}+s+t-i
\end{array} \bigg|1\right] N^{st}_{i}(m,n), \end{equation}
$q=\frac{1}{4}$ and: 
\begin{equation}
\label{eq:NStr}
N^{st}_{i}(m,n)=\scriptstyle\sum_{k=0}^{i-1}(-1)^{k} \begin{pmatrix}i-1\\ k\end{pmatrix} \bigl(s-1+m+1\bigr)_{k-i+1} \bigl(s-1-m+1\bigr)_{-k} \bigl(t-1+n+1\bigr)_{-k} \bigl(t-1-n+1\bigr)_{k-i+1}.
\end{equation}
Where the $(n)_{k}$ are the ascending Pochhammer symbols. 

Be our definition of trace
\begin{equation}
tr\left(V^s_{m_s}V^s_{-m_s}\right)\equiv\frac{6}{1-\lambda^2}   \frac{(-1)^{m_s} 2^{3-2s}\Gamma(s+m_s)\Gamma(s-m_s)}{(2s-1)!!(2s-3)!!} \prod^{s-1}_{\sigma=1}\left(\lambda^2-\sigma^2\right) \label{trace}
\end{equation}
In this  paper we take $\lambda=3$ and remain with the ideal part, $2\leq s \leq3$.\\

The Killing metric on the principal embedding for the ordering given in (\ref{Ppalemb})
\begin{equation}
g_{ab}=tr(V_a V_b)=\left(
\begin{array}{cccccccc}
 0 & 0 & 1 & 0 & 0 & 0 & 0 & 0 \\
 0 & -\frac{1}{2} & 0 & 0 & 0 & 0 & 0 & 0 \\
 1 & 0 & 0 & 0 & 0 & 0 & 0 & 0 \\
 0 & 0 & 0 & 0 & 0 & 0 & 0 & -1 \\
 0 & 0 & 0 & 0 & 0 & 0 & \frac{1}{4} & 0 \\
 0 & 0 & 0 & 0 & 0 & -\frac{1}{6} & 0 & 0 \\
 0 & 0 & 0 & 0 & \frac{1}{4} & 0 & 0 & 0 \\
 0 & 0 & 0 & -1 & 0 & 0 & 0 & 0
\end{array}
\right)
\end{equation}\\

The Killing metric in diagonal embedding for the ordering given in (\ref{basisW32N})

\begin{equation}
g_{ab}=tr(V_aV_b)=\left(
\begin{array}{cccccccc}
 -\frac{1}{8} & 0 & 0 & 0 & 0 & 0 & 0 & 0 \\
 0 & 0 & \frac{1}{4} & 0 & 0 & 0 & 0 & 0 \\
 0 & \frac{1}{4} & 0 & 0 & 0 & 0 & 0 & 0 \\
 0 & 0 & 0 & -\frac{1}{6} & 0 & 0 & 0 & 0 \\
 0 & 0 & 0 & 0 & 0 & 0 & -\frac{1}{4} & 0 \\
 0 & 0 & 0 & 0 & 0 & 0 & 0 & -\frac{1}{4} \\
 0 & 0 & 0 & 0 & -\frac{1}{4} & 0 & 0 & 0 \\
 0 & 0 & 0 & 0 & 0 & -\frac{1}{4} & 0 & 0
\end{array}
\right)\end{equation}
\\

\paragraph{Useful results}

Here we report some results that were useful during the computations in section \ref{subsec:RT}.
In particular, the solution to the conditions
 \begin{eqnarray}
 (\delta \mathcal{L}^{(0)}_1)_{\delta \rightarrow(\delta_{\Lambda})|_{\mu_3\rightarrow 0}}&=& (\delta_{\Lambda}\mathcal{L})\bigg|_{\text{ At }\mu_3~ \&~x_2\rightarrow0},\nonumber \\
(\delta \mathcal{W}^{(0)}_1)_{\delta \rightarrow(\delta_{\Lambda})|_{\mu_3\rightarrow 0}}&=& (\delta_{\Lambda}\mathcal{W})\bigg|_{\text{ At }\mu_3~ \&~x_2\rightarrow0}, \label{Consistency}
 \end{eqnarray}
 where we remind the reader that by $(\delta \ldots)|_{\delta\rightarrow\delta_\Lambda}$ we mean:
 \begin{itemize}
 
\item  Take the functional differential of $\ldots$ in terms of $(\delta \mathcal{L}^{(0)},\delta \mathcal{W}^{(0)})$ and therafter substitute $\delta$ by $\delta_{\Lambda}$. The expressions for $(\delta_{\Lambda} \mathcal{L}^{(0)},\delta_{\Lambda} \mathcal{W}^{(0)})$ are reported in (\ref{deltalambda}). The expressions for $(\delta_{\Lambda} \mathcal{L},\delta_{\Lambda} \mathcal{W})$ are  reported in (\ref{gaugeVARIATIONS}).

\end{itemize}

The most general solution to (\ref{Consistency}) read out
\begin{eqnarray}
\mathcal{L}^{(0)}_1&=&3 c_1\mathcal{W}^{(0)}+c_2\partial_{1}  \mathcal{L}^{(0)}+2 c_1 x_1
 \partial_{1}  \mathcal{W}^{(0)} \nonumber\\
\mathcal{W}^{(0)}_1&=&-c_1\left(\frac{8}{3}{\mathcal{L}^{(0)}}^2+ \frac{3}{4}\partial_{1} ^2\mathcal{L}^{(0)}\right)+c_2\partial_{1}  \mathcal{W}^{(0)}- c_1 x_1\left(
\frac{8}{3}\partial_{1} \mathcal{L}^{(0)}+\frac{1}{6}\partial_{1} ^3\mathcal{L}^{(0)}\right)\nonumber\\\epsilon^{(0)}_1&=&-c_1 \left(\frac{8}{3}\eta^{(0)} \mathcal{L}^{(0)}+\frac{1}{4}\partial_{1} ^2 \eta^{(0)}\right) +c_2 \partial_{1}  \epsilon^{(0)}+ c_1 x_1\left(\frac{8}{3}\partial_{1}  \eta^{(0)} \mathcal{L}^{(0)}+ \frac{1}{6}\partial_{1}^3\eta^{(0)}\right)\nonumber \\
\eta^{(0)}_1&=& c_1 \epsilon^{(0)} +c_2 \partial_{1}  \eta^{(0)}-2 c_1 x_1 \partial_{1} \epsilon^{(0)}. \label{Intconst0}
\end{eqnarray}
It is straightforward to check that (\ref{Intconst0}) coincides with (\ref{RedefPar}) for $c_1=1$ and $c_2=0$. In fact this is the unique choice out of (\ref{Intconst0}) that allows to integrate the differential of charge to (\ref{chargeW3}).

It is also useful to write down the most general choice of $(\mathcal{L}^{(0)}_1,\mathcal{W}^{(0)}_1,\epsilon^{(0)}_1,\eta^{(0)}_1)$ that is consistent without explicit dependence on $\phi$ and dimensional analysis. It is given by
\begin{eqnarray}
{\mathcal{L}^{(0)}}_{1 hom}&=&c_3 \mathcal{W}+c_4 \partial_{1} \mathcal{L},~~~~~~~~~~ {\mathcal{W}^{(0)}}_{1 hom}=c_5 \mathcal{L}^2+c_6\partial_{1} ^2 \mathcal{L}+c_7\partial_{1} \mathcal{W},\nonumber\\ {\epsilon^{(0)}_{1}}_{hom}&=&c_8 \partial_{1} \epsilon+c_9 \partial_{1} ^2 \eta+2 c_{10} \mathcal{L}\eta,~ ~~~~{\eta^{(0)}_1}_{hom}= c_{11} \epsilon+c_{12} \partial_{1}  \eta. \label{Ansatz}
\end{eqnarray}\\
We use (\ref{Ansatz}) to show that (\ref{Dbras3}) is not isomorphic to $W_3$.

\bibliographystyle{JHEP}
\bibliography{sl3}
\end{document}